\documentclass[12pt,reqno]{amsart}
\usepackage{comment}
\usepackage{MnSymbol}  
\usepackage{amsmath}
\usepackage{amsfonts}
\usepackage{bm}
\usepackage{array}  
\usepackage{hyperref}
\usepackage{enumerate}
\usepackage{tikz}
\usepackage{graphicx}
\usepackage{tikz-cd}
\usetikzlibrary{calc,positioning}
\tikzset{
wei/.style={circle,fill, minimum size=0.4pt,inner sep=0.8pt},
}
\newcommand\bit{\begin{itemize}}
\newcommand\eit{\end{itemize}}
\newcommand\lan{\langle}
\newcommand\ran{\rangle}
\newcommand{\bp}{\begin{pmatrix}}
\newcommand{\ep}{\end{pmatrix}}
\newcommand\Pa{Painlev\'e }
\newcommand{\mb}{\mathbb}

\newcommand{\oc}[1]{{#1}^{\vee}}
\usepackage{setspace}
\usepackage{fullpage}
\usepackage{marginnote}
\definecolor{mycol}{rgb}{0,0.5,0}

\newcommand{\al}{\alpha}

\newcommand{\Om}{\Omega}
\newcommand{\de}{\delta}

\newcommand{\De}{\Delta}
\newcommand{\be}{\beta}

\newcommand{\ga}{\gamma}

\newcommand{\s}{\sigma}

\theoremstyle{break}    
\newtheorem{Rem}{Remark} \theoremstyle{marginbreak}

\newtheorem{eg}{Example}
\newcommand{\beqn}{\begin{equation}}
\newcommand{\eeqn}{\end{equation}}

\begin{document}
\author{Yang Shi}
\address{College of science and engineering, Flinders at Tonsley, Flinders University, SA 5042,
    Australia}
\email{yang.shi@flinders.edu.au}
\title{Two variations on $(A_3\times A_1\times A_1)^{(1)}$ type discrete Painlev\'e equations}
\maketitle
\begin{abstract}
By considering the normalizers of reflection subgroups of types $A_1^{(1)}$ and $A_3^{(1)}$ in $\widetilde{W}\left(D_5^{(1)}\right)$,
two normalizers: $\widetilde{W}\left(A_3\times A_1\right)^{(1)}\ltimes {W}(A_1^{(1)})$
and $\widetilde{W}\left(A_1\times A_1\right)^{(1)}\ltimes {W}(A_3^{(1)})$ can be
constructed from
a $(A_{3}\times A_1\times A_1)^{(1)}$ type subroot system.
These two symmetries arose in the studies of discrete \Pa equations \cite{KNY:2002, Takenawa:03, OS:18},
where certain non-translational elements of infinite order
were shown to give rise to discrete \Pa equations.
We clarify the nature of these elements
in terms of Brink-Howlett theory of normalizers of Coxeter groups \cite{BH}.
This is the first of a series of studies which investigates the properties of
discrete integrable equations 
via the theory of normalizers.
\end{abstract}
\section{Introduction}
Formulation of \Pa equations and their generalisations as birational representations
of Weyl groups provides us with an elegant and efficient way to characterise and study these highly transcendental, nonlinear equations. In particular,
it is well-known that discrete evolutions of \Pa equations are given by 
translational elements of extended affine Weyl groups. Examples include the work of Okamoto on the symmetries of the six classical \Pa equations \cite{oka:79},
Sakai's classification of second-order discrete \Pa equations \cite{sak:01}, and  
generalisations of \Pa equations such as Kajiwara, Noumi and Yamada's
$\widetilde{W}\left(A_n\times A_m\right)^{(1)}$ system (KNY) \cite{KNY:2002}, Sasano \cite{sa:06} and Masuda's \cite{Masuda:15} $D_n$ type systems, and more recently Okubo and Suzuki's 
$(A_{2n+1}\times A_1\times A_1)^{(1)}$ system \cite{OS:18}.

Theory of Weyl groups, or in general Coxeter groups \cite{Hbook} is a classical area of algebra
rich with remarkable properties, for which significant results and
breakthroughs are still being made, long-standing conjectures proved \cite{BW:14}.
Characterisation of integrable systems in terms of such groups
naturally leads us to the question: how much of the intrinsic properties of the group
remain in the realisations of 
integrable equations and what are their implications in this particular context?
For example, varieties of subgroup relations exist between Coxeter groups of different type.
Do different integrable systems relate in a way that is inherent with these subgroup structures?

An example of this kind was given by 
an interesting link discovered by Takenawa \cite{Takenawa:03}  
between a KNY system for the case $n=1, m=3$, that is a $\widetilde{W}\left(A_1\times A_3\right)^{(1)}$ type discrete \Pa
equation and Sakai's $\widetilde{W}\left(D_5^{(1)}\right)$ $q$-\Pa equation.
It was thought that KNY's $\widetilde{W}\left(A_1\times A_3\right)^{(1)}$ system should be a second-order $q$-differece \Pa equation, however it does not coincide in symmetry with any equations in Sakai's list. 
In 2003 Takenawa 
established via algebraic geometrical means, the fact that KNY's $\widetilde{W}\left(A_1\times A_3\right)^{(1)}$ equation can be embedded as a sub-system of Sakai's $\widetilde{W}\left(D_5^{(1)}\right)$ equation. In particular, it was found
that in this embedding the element $\phi$ which give rise to $\widetilde{W}\left(A_1\times A_3\right)^{(1)}$ type \Pa equation is not a translation in $\widetilde{W}\left(D_5^{(1)}\right)$, whereas the element $\phi^2$ is translational. Similar examples have been found in different contexts in the integrable system literature
\cite{KNT:11, Stokes:18,ahjn:16,cp:17}. A common feature of these examples is that the element $\phi$ which gives rise to
a discrete \Pa equation is not translational, whereas some powers of $\phi$ is. We refer to
such an element as {\it quasi-translational}. In the present work,
we show that quasi-translational
elements are in fact certain elements of infinite order occur in normalizers of Weyl groups. Moreover, we clarify the quasi-translational nature of such elements in terms
of normalizer theory.

Our approach is illustrated through a simple example. We look at
the normalizer theory for an underlying 
$(A_3\times A_1\times A_1)^{(1)}$ type subsystem
of a $D_5^{(1)}$ root system.
We show that Takenawa's embedding
of KNY's $\widetilde{W}\left(A_1\times A_3\right)^{(1)}$ equation and
the $n=1$ case of Okubo and Suzuki's 
$(A_{3}\times A_1\times A_1)^{(1)}$ system coincide with the two variations of
normalizers arising from an $(A_{3}\times A_1\times A_1)^{(1)}$ subsystem
of Sakai's $\widetilde{W}\left(D_5^{(1)}\right)$ $q$-\Pa equation.

The paper is organised as follows.
In Section \ref{SCox} we give a brief summary of some well-known facts and properties of Coxeter groups relevant 
to our discussion. 
The main results are given in Section \ref{SCND5},
where we show that two  
subgroups of $\widetilde{W}\left(D_5^{(1)}\right)$ with
an underlying 
$(A_3\times A_1\times A_1)^{(1)}$ subroot system
: $\widetilde{W}\left(A_3\times A_1\right)^{(1)}\ltimes {W}(A_1^{(1)})$ and $\widetilde{W}\left(A_1\times A_1\right)^{(1)}\ltimes {W}(A_3^{(1)})$ arise as the
normalizers of Weyl subgroups of type $A_1^{(1)}$ and $A_3^{(1)}$, respectively.
We define quasi-translational elements as certain elements of 
the normalizer, their properties are described in detail in Sections \ref{Squasi} and \ref{2var}.
In Section \ref{TOS}, the normalizer theory
developed in Section \ref{SCND5} is placed in the
context of discrete \Pa equations. In particular, we discuss the quasi-translational nature of the elements which give rise to discrete \Pa equations given by Takenawa \cite{Takenawa:03}, and Okubo and Suzuki \cite{OS:18} in terms of
the normalizer theory. Concluding remarks and some future directions
are given in Section \ref{Dis}.

\section{Properties of Coxeter groups}\label{SCox}
Let $W=\langle s_i\mid s_i^2=1, (s_is_j)^{m_{ij}}=1, \; 1\leq i,j \leq n\rangle$
be a finite {\it reflection group} or {\it Coxeter
group} whose defining relations of the generators are encoded in a corresponding
Dynkin diagram. When parameter $m_{ij}\in\{2,3,4,6\}$,
known as the {\it crystallographic condition}, W is called a finite Weyl group. Its affine extension 
$W^{(1)}=\langle s_i\mid s_i^2=1, (s_is_j)^{m_{ij}}=1, \; 0\leq i,j \leq n\rangle$,
is a group of infinite order
with a corresponding extended Dynkin diagram.
Dynkin diagram is a diagram consisting of vertices and bonds: each vertex of the diagram represents a generator $s_i$ is labeled
by $i$, for $0\leq i\leq n$. The parameter $m_{ij}$ takes value of: $2, 3, 4$ or $6$ 
when two vertices labeled $i$ and $j$
are respectively: disconneted, joined by a single, a double, or a triple bond. Diagrams which have only
single bonds are called {\it simply-laced}, they are of types $A_n$, $D_n$, $E_6$, $E_7$ and $E_8$. 
The non-simply laced types are $B_n$, $C_n$, $F_4$ and $G_2$. 
For each $s_i$ in the generating set of $W^{(1)}$, known as a {\it simple reflection},  we have a
corresponding {\it simple root} $\al_i$. Vertices of Dynkin diagram
can be equivalently labelled by either simple reflections or simple roots.
We have 
\beqn
\De=\{ \al_i\mid 1\leq i \leq n\}\quad\mbox{ and }\quad \De^{(1)}=\{ 
\al_i\mid 0\leq i \leq n\},
\eeqn which are the {\it simple systems} of finite and affine type, respectively.
The finite and affine {\it root systems} are then given respectively by 
\beqn
\Phi=W.\De\quad\mbox{ and }\quad\Phi^{(1)}=W^{(1)}.\De^{(1)}.
\eeqn
Any root of  
$\Phi^{(1)}$ can be expressed in the form
$\al+k \de$ for $k\in \mathbb{Z}$ and $\al\in \Phi$, where
\beqn\label{de}
\de=\al_{0}+\tilde{\al}=\al_{0}+\sum_{i=1}^{n}c_i\al_i,
\eeqn
is called the {\it null root}, and $\tilde{\al}$ the {\it highest root}.
The value of $c_i$ for all Weyl groups can be found in 
any reference books on Coxeter groups, \cite{Hbook}
for example. 
{\eg\label{ED5} A Weyl group of $D^{(1)}_5$ type.
Let $W(D_5)=\langle s_i\mid 1\leq i \leq 5\rangle$ and
$W(D^{(1)}_5)=\langle s_i\mid 0\leq i \leq 5\rangle$ be the finite and affine Weyl group of type $D_5$,
respectively. 
Their defining relations are encoded in the Dynkin diagram of Figure \ref{D5}.
\begin{figure}[ht!]
\begin{tikzpicture}[thick]
\node  (a1) {};
\node [right=of a1, wei](a2){,};
\node [right=of a2, wei](a3) {.};
\node [right=of a3](a5){};
\node [above=of a5, wei](a8) {.};
\node [below=of a5, wei](a9) {.};
\node [above=of a1, wei](a10) {.};
\node [below=of a1, wei](a11) {.};
%
\node (r1) at ($(a2)!0.5!(a3)$) {};
\node [above=of r1](a12) {};

\draw (a3) node [anchor=north] {$3$} ;
\draw (a8) node [anchor=west] {$5$} ;
\draw (a9) node [anchor=west] {$4$} ;
\draw (a2) node [anchor=north] {$2$} ;
\draw (a10) node [anchor=east] {$0$} ;
\draw (a11) node [anchor=east] {$1$} ;
\draw (a1) node [anchor=east] {$\s_1$} ;
\draw (a1) node {$\curvearrowupdown$};
\draw (a12) node [anchor=south] {$\s_2$};
\draw (a12) node[rotate=-90] {$\curvearrowupdown$};

\draw[-] (a3) -- (a8);
\draw[-] (a3) -- (a9);
\draw[-] (a2) --  (a10);
\draw[-] (a2) -- (a11);
\draw[-] (a2) -- (a3);
\end{tikzpicture}
\caption{Dynkin diagram of type ${D}^{(1)}_{5}$}\label{D5}
\end{figure}
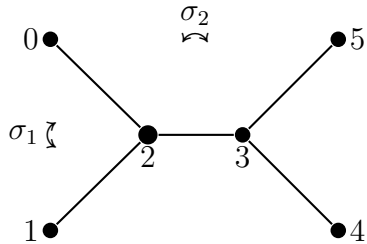

Simple systems of finite and affine type
are given respectively by $\De=\{ \al_i\mid 1\leq i \leq 5\}$ and $\De^{(1)}=\{ 
\al_i\mid 0\leq i \leq 5\}$; $\Phi=W({D}_{5}).\De$ and 
$\Phi^{(1)}=W({D}^{(1)}_{5}).\De^{(1)}$ 
the finite and affine root systems. The null root in this case is given by
\beqn\label{deD5}
\de=\al_{0}+\tilde{\al}=\al_0+\al_{1}+2\al_{2}+2\al_{3}+\al_{4}+\al_{5}.
\eeqn 
From now on
we adapt to the notation: $\al_i+...+\al_j=\al_{i...j}$ to
express a sum of simple roots,
and $s_i...s_j=s_{i...j}$ for product of simple reflections.
For example we now write $\de=\al_{01223345}$. }

Finite and affine Weyl groups $W$ and $W^{(1)}$ can be realised as groups of reflections in real vector spaces $V$ and $V^{(1)}$ with bases $\De$ and $\De^{(1)}$, respectively. From definition of $\de$, we see that the set 
$\{ \al_1, \ldots, \al_n,\de\}$ is also a basis of $V^{(1)}$.
For our purpose, it is
instructive to introduce the dual spaces of $V$ and $V^{(1)}$, given by the bilinear pairing
$\langle {}\,, {} \rangle:$ $V^{(1)}\times V^{(1)\ast} \to \mathbb R$,
\begin{equation}\label{w}
\langle \al_i, h_j\rangle=\delta_{ij}, \quad\langle \al_i, h_{\de}\rangle=
\langle \de, h_{j}\rangle=0\;\mbox{for}\; (1\leq i, j\leq n),\quad\mbox{and}\quad\langle \de,h_\de \rangle=1.
\end{equation} 
Dual spaces $V^{\ast}$ and $V^{(1)\ast}$ have bases 
$\{ h_1, \ldots, h_n\}$ and $\{ h_1, \ldots, h_n, h_\de\}$, respectively.
Vectors $h_i\;(1\leq i\leq n)$ are called the {\it fundamental weights}, and
$P=\mathbb Z \{ h_1, \ldots, h_n\}$ is the
{\it weight lattice}.

Simple coroots $\oc \al_i\in V^{(1)\ast}$
are defined by
\begin{equation}\label{Cartan}
\langle\al_i,\oc \al_j\rangle=A_{ij}
\end{equation}
for all $0\leq i, j\leq n$, and $(A_{ij})_{i,j=0}^n$ is the generalised Cartan matrix.
Let $Q=\mathbb Z \{ \oc\al_1, \ldots, \oc\al_n\}$ be the coroot lattice.
It can be easily checked that as a consequence of Equations \eqref{w} and \eqref{Cartan} we have
\begin{equation}\label{rw}
\oc\al_i=\sum_{j=1}^{n}A_{ij}h_j,
\end{equation}
and in particular for simply-laced Weyl groups we have
\begin{equation}\label{rw0}
\oc\al_0
=-\sum_{i=1}^{n}c_i\oc\al_i,
\end{equation}
where $c_i$ are given in Equation \eqref{de}.
Let $s_\al$ denotes
reflection along the hyperplane orthogonal to a root $\al\in \Phi^{(1)}$.
The generator $s_i=s_{\al_i}$ of $W^{(1)}$ then is reflection along the hyperplane
orthogonal to the simple root $\al_i\in\De^{(1)}$, given by the formula,
\begin{equation}\label{siC}
s_i\al_j=\al_j-\langle\al_j,\oc \al_i\rangle\al_i=\al_j-A_{ji}\al_i.
\end{equation}
$W^{(1)}$ acts on $V^{(1)\ast}$ via the contragredient action:
\beqn\label{cong}
\lan w^{-1} f,  h\ran=\lan f, wh\ran, \quad\mbox{for}\quad f\in V^{(1)}, h\in V^{(1)\ast}, w\in W^{(1)}.
\eeqn
That is $s_i$ acts on the dual space $V^{(1)\ast}$ by ,
\begin{equation}\label{siCi}
s_{i} \oc\al_j=\oc \al_j-A_{ij}\oc \al_i,
\end{equation}
for all $0\leq i, j\leq n$.
Moreover we have the following
property for elements of reflection
\beqn\label{asa}
s_{w(\al)}=ws_\al w^{-1},\quad \mbox{for}\quad\al\in \De^{(1)}, \quad w\in W^{(1)}.
\eeqn 
The group $W^{(1)}$ acts transitively
on the root system, that is for all $\be_1, \be_2\in \Phi^{(1)}$, there is
a $w\in W^{(1)}$ such that $w\be_1=\be_2$, which means that reflection associated to any root of the root system can be presented as a conjugation to a simple reflection.
{\eg Weyl group of $D_5^{(1)}$ type realised as group of reflections.
Real vector spaces $V$
and $V^{(1)}$ for which the Weyl group of $D^{(1)}_5$ type can be realised as
groups of reflections
have bases $\De=\{ \al_i\mid 1\leq i \leq 5\}$ and $\De^{(1)}=\{ 
\al_i\mid 0\leq i \leq 5\}$, respectively.
The corresponding dual spaces $V^{\ast}$ and $V^{(1)\ast}$ with bases $\{\, h_1, h_2, h_3, h_4, h_5\}$ and $\{\, h_1, h_2, h_3, h_4, h_5, h_\de\}$ are defined by the bilinear pairing
$\langle {}\,, {} \rangle:$ $V^{(1)}\times V^{(1)\ast} \to \mathbb R$,
\begin{equation}\label{wD5}
\langle \al_i, h_j\rangle=\delta_{ij}, \quad\langle \al_i, h_{\de}\rangle=
\langle \de, h_{j}\rangle=0\;\mbox{for}\; (1\leq i, j\leq 5),\quad\mbox{and}\quad\langle \de,h_\de \rangle=1.
\end{equation}
Vectors $h_i\;(1\leq i\leq 5)$ are the fundamental weights, and
$P=\mathbb Z \{h_1, h_2,h_3, h_4,h_5\}$
is the $D_5$ weight lattice. 
The set of simple coroots in $V^{(1)\ast}$ is
$\oc {\De^{(1)}}=\{\oc\al_0, \oc\al_1,  \oc\al_2,\oc\al_3, \oc\al_4, \oc\al_5\}$
where we have
\begin{equation}\label{CartanD5}
\langle\al_i,\oc \al_j\rangle=A_{ij}
\end{equation}
for all $0\leq i, j\leq 5$, and $(A_{ij})_{i,j=0}^5$ is the generalised Cartan matrix of type 
$D_5$ given by
 \begin{equation}
 (A_{ij})_{i,j=0}^5=\begin{pmatrix}2&0&-1&0&0&0\\0&2&-1&0&0&0\\-1&-1&2&-1&0&0\\
 0&0&-1&2&-1&-1\\
 0&0&0&-1&2&0\\0&0&0&-1&0&2\end{pmatrix}.
\end{equation}
Moreover we have
\begin{equation}\label{rwD5}
\oc\al_i=\sum_{j=1}^{5}A_{ij}h_j\quad \mbox{and}\quad \oc\al_0
=-(\oc\al_1+2\oc\al_2+2\oc\al_3+\oc\al_4+\oc\al_5)= -h_2.
\end{equation}

The group $W({D}^{(1)}_{5})$ is realised as a group of reflections in $V^{(1)}$ and $V^{(1)\ast}$
by formulae \eqref{siC} and \eqref{siCi}, respectively.
}
\subsection{Translations in the weight lattice.}\label{STran}
One can form an
{\it extended affine Weyl group} $\widetilde{W}^{(1)}=W^{(1)}\rtimes A$, where
$A$ is a group of certain Dynkin diagram automorphisms
acting on $W^{(1)}$ via conjugation. 
It is a remarkable property of the Weyl group that with appropriate
extension $\widetilde{W}^{(1)}$
decomposes into
a semidirect product of the finite Weyl group $W$ and translations on weight lattice $P$:
$\widetilde{W}^{(1)}=W\ltimes P$.
Translational elements of $\widetilde{W}^{(1)}$ are best understood by looking
at their actions on
a hyperplane $H$ in the dual space $V^{(1)\ast}$ defined by,
\begin{equation}\label{H}
H=\{\,h\in V^{(1)\ast}\,|\, \langle \de, h\rangle=1\},\quad\mbox{for}\quad h\in H.
\end{equation}
Let $\oc\mu\in V^{\ast}$ be a point on the weight lattice $P$, that is $\oc\mu=\sum_{i=1}^{n}\mu_{i}h_{i}$, for $\mu_i\in \mathbb{Z}$. We define a translational element $t_{\oc\mu}\in \widetilde{W}^{(1)}$
such that it acts on $h\in H$ by
\beqn\label{Transn}
t_{\oc\mu} h= h+\oc\mu,
\eeqn
whereas its action on simple affine roots $\al_{i} \in \De^{(1)}$ is given by
 \beqn\label{Transl}
 t_{\oc\mu}\al_i=\al_i-\lan\al_i, \oc\mu\ran\de=\al_i-\mu_i \de.
 \eeqn
That is, $t_{\oc\mu}$ shifts $\al_i$ by $-\mu_i$ multiples of $\de$ for $1\leq i\leq n$.
By 
property of the null root given in Equation \eqref{w}, coefficients $\mu_i$ ($0\leq i\leq n$) satisfy the constraint
\begin{align}\label{Tranc}
  0=\lan\de, \oc\mu\ran&=\lan\sum_{i=0}^{n}c_{i}\al_{i}, \oc\mu\ran=\sum_{i=0}^{n}c_{i}\mu_i,
\end{align}
where $c_0=1$ .
Moreover, we have the following property for translational elements,
\beqn\label{cj}
wt_{\oc\mu} w^{-1}=t_{w\oc\mu},\quad w\in \widetilde{W}^{(1)}.
\eeqn
Let the column vector
$(a_1, \ldots, a_n, a_\de)^{T}$ where $a_1, \ldots, a_n, a_\de\in \mathbb{R}$  be the coordinate vector of the dual space $V^{(1)\ast}$
in basis $\{h_1,\ldots, h_n, h_{\de}\}$.
Then by definition given in Equation \eqref{H}, all vectors in $H\subset V^{(1)\ast}$
are of the form $(a_1, \ldots, a_n, 1)^{T}$. Translational element 
$t_{\oc\mu}$ as an $(n+1)\times(n+1)$ matrix of linear transformation on $V^{(1)\ast}$ in basis $\{h_1,\ldots, h_n, h_{\de}\}$ acting from the left
given by Equation \eqref{Transn} is then:
\beqn
t_{\oc\mu}^{h}=
\bp
I_n & \bm{\oc\mu}\\
0 & 1
\ep,
\eeqn
where $I_n$ is the $n\times n$ Identity matrix, and 
$\bm{\oc\mu}=(\mu_1, ..., \mu_n)^T$ denotes the $n$-column 
coordinate vector of $\oc\mu$.
By contragredient action element 
$t_{\oc\mu}$ as an $(n+1)\times(n+1)$ matrix of linear transformation on $V^{(1)}$ in  basis $(\al_1, \ldots, \al_n, \de)$ acting from the right is
given by Equation \eqref{Transl}: 
\beqn
t_{\oc\mu}^{\de}=(t_{\oc\mu}^{h})^{-1}=
\bp
I_n & -\bm{\oc\mu}\\
0 & 1
\ep.
\eeqn
{\eg Translations in $\widetilde{W}\left(D_5^{(1)}\right).$

Let $\s_1, \s_2$ be two diagram automorphisms of $D_5^{(1)}$ type Dynkin diagram (see Figure \ref{D5}),
acting on the simple reflections
of $W(D^{(1)}_5)$ via conjugation:
\begin{align*}
\s_1s_{\{0,1,2,3,4,5\}}&=s_{\{1,0,2,3,4,5\}}\s_1, \\
\s_2s_{\{0,1,2,3,4,5\}}&=s_{\{5,4,3,2,1,0\}}\s_2, \\
\s_1\s_2s_{\{0,1,2,3,4,5\}}&=s_{\{5,4,3,2,0,1\}}\s_1\s_2. 
\end{align*}
If we write the above actions as permutations of the index set $\{0,1,2,3,4,5\}$ we have:
\[
\s_1=(10),\quad
\s_2=(05)(14)(23),\quad
\s_1\s_2=(5140)(23).
\]
Then it is easily seen that element $\s_1\s_2$ is of order 4. The fact that 
the index of connection between the weight and coroot lattice for $D_{odd}$  type
Weyl groups  is 4 tells us that we need to extend $W(D^{(1)}_5)$ by $A=\lan \s_1\s_2\ran$,
a cyclic group of order four, in order for the
extended affine Weyl group to
have the decomposition of finite Weyl group and weight lattice $P$, that is $\widetilde{W}\bigl(D^{(1)}_5\bigr)=W(D^{(1)}_5)\rtimes A=W(D_5)\ltimes P$.
}

In the following, we give explicit examples of
elements of translation in $\widetilde{W}\bigl(A^{(1)}_1\bigr)$ and
$\widetilde{W}\bigl(A^{(1)}_3\bigr)$ which will be relevant for our later discussion of 
quasi-translations in 
the context of discrete \Pa equations.

{\eg\label{TranA1} Translations in $\widetilde{W}\bigl(A^{(1)}_1\bigr)$.

For a simple system $\De^{(1)}=\{\al_1, \al_0\}\subset V^{(1)}$ of type $A^{(1)}_1$ , we have
the extended affine Weyl group $\widetilde{W}\bigl(A^{(1)}_1\bigr)
=\langle s_1, s_0, \pi\rangle$,
where $\pi$ is the Dynkin diagram automorphism
that exchanges the two roots $\al_1$ and $\al_0$. 
Defining relations of the generators are encoded by the Dynkin diagram in Figure \ref{CDA1}.
\begin{figure}[!htbp]
\begin{tikzpicture}[scale=1]
\node  (a1) {$\circ$};
\node [right=of a1](a2) {} ;
\node [right=of a2](a3) {$\circ$} ;
\filldraw [black](a1) node [anchor=north] {$s_1$}circle(.5ex) ;
\draw (a2) node [anchor=north] {$\infty$} ;

\filldraw [black] (a3) node [anchor=north] {$s_0$}circle(.5ex) ;
\draw[-] (a1)--(a3);
\node (r1) at ($(a1)!0.5!(a3)$) {};
\draw (r1) node [anchor=south] {$\pi$} ;
\draw (1.4, 0.5) node  [rotate=-90] {$\curvearrowupdown$};
\path[use as bounding box] (-1.5,0) rectangle (0,0);
\end{tikzpicture}
\caption{Dynkin diagram of $\widetilde{W}\bigl(A^{(1)}_1\bigr)$}\label{CDA1}
\end{figure}
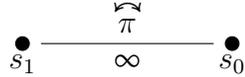

The null root in this case is given by
\beqn\label{nrA1}
\de=\al_0+\al_1.
\eeqn

The fundamental weight $h_1\in V^{(1)\ast}$ is defined
by 
\beqn\label{A1h1}
\langle \al_1,h_1\rangle=1, \quad \mbox{and} \quad \langle \de,h_1\rangle=0.
\eeqn
This together with the definition of the null root in Equation \eqref{nrA1}
 implies that $\langle \al_0,h_1\rangle=-1$.
Coroot $\oc\al_1$ is related to $h_1$ by the Cartan matrix of type $A_1$,
\beqn
\oc\al_1=2h_1.
\eeqn
The weight lattice is $P=\mathbb{Z}\{h_1\}$.
Translation on the weight lattice by $h_1$ is given by the element  
\beqn\label{th1}
t_{h_1}=\pi s_1,
\eeqn
which as matrices of linear transforms on vector spaces $V^{(1)\ast}$ and $V^{(1)}$ in bases $\{h_1, h_{\de}\}$ and $\{\al_1, \de\}$
are respectively:
\beqn
t_{h_1}^{h}=
\bp
1 & 1\\
0 & 1
\ep\quad\mbox{and}\quad
t_{h_1}^{\de}=
\bp
1 & -1\\
0 & 1
\ep.
\eeqn
For example,  we can read off the action of element $t_{h_1}$ on root
$\al_1$ by looking at the first row of the matrix $t_{h_1}^{\de}$ , that is we have
\beqn
t_{h_1} : \al_1 \mapsto \al_1-\de.
\eeqn
The action of $t_{h_1}$ on $\al_0$ can then  be inferred by definition \eqref{nrA1} of $\de$ and the condition given in 
Equation \eqref{Tranc}, hence we have 
\beqn\label{th1l}
t_{h_1} : \{\al_1, \al_0\} \mapsto \{\al_1-\de, \al_0+\de\}.
\eeqn
}
{\eg\label{TranA3} Translations in $\widetilde{W}\bigl(A^{(1)}_3\bigr)$.

For a simple system $\De^{(1)}=\{\be_0, \be_1, \be_2, \be_3\}$ of type $A^{(1)}_3$, we have
$\widetilde{W}\bigl(A^{(1)}_3\bigr)
=\lan s_{\be_0}, s_{\be_1}, s_{\be_2}, s_{\be_3},
p_1p_2\ran$, whose defining relations are encoded in the Dynkin diagram of Figure \ref{DynA3}.
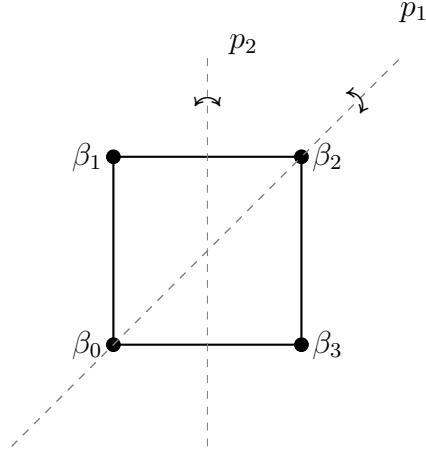
\begin{figure}[t]
\begin{center}
%
\begin{tikzpicture}[scale=.5]
\filldraw [black]  (2,2) node [anchor=east] {$\be_0$} circle(1ex);
\filldraw [black] (2,7) node [anchor=east] {$\be_1$} circle(1ex);
\filldraw [black]  (7,2) node [anchor=west]{$\be_3$} circle(1ex);
\filldraw [black]  (7,7) node [anchor=west] {$\be_{2}$ } circle(1ex);
\draw [black]  (4.5,2) node [anchor=north] {} ;
\draw [black]  (4.5,7) node [anchor=south] {} ;
\draw [black]  (2,4.5) node [anchor=east] {} ;
\draw [black]  (7,4.5) node [anchor=west] { } ;

\draw [thick] (2,7) -- (7,7);
\draw [thick] (7,2) -- (7,7);
\draw [thick](2,2) -- (2,7);
\draw [thick] (2,2) -- (7,2);
\node (sw) at (-1,-1) {}; 
\node (w) at (-1,4.5) {}; 
\node (s) at (4.5,-1) {}; 
\node (n) at (4.5,10) [label=0:$p_{2}$] {}; 
\node (se) at (10,-1) {}; 
\node (ne) at (10, 10) [label=90:$p_{1}$]{}; 
\draw[dashed, gray] (sw)--(ne);
\draw[dashed, gray] (sw)--(ne);
\draw[dashed, gray] (s)--(n);
\draw (8.5, 8.5) node [rotate=-145] {$\curvearrowupdown$};
\draw (4.5, 8.5) node [rotate=-90] {$\curvearrowupdown$};
\end{tikzpicture}
%
%
%
\caption{Dynkin diagrams of 
$\widetilde{W}\bigl(A^{(1)}_3\bigr)$ system}\label{DynA3}
\label{DynA3}
\end{center}
\end{figure}
Elements $p_1$ and $p_2$ are diagram automorphisms
of order 2,
\begin{align*}
p_1&: \{\be_1\leftrightarrow\be_3\},\\
p_2&: \{\be_1\leftrightarrow\be_2,\be_0\leftrightarrow\be_3\},
\end{align*}
whereas the element $p_1p_2$ is 
of order 4,
\beqn p_1p_2:\{\be_0, \be_1, \be_2, \be_3\} \mapsto\{\be_1, \be_2, \be_3, \be_0\},
\eeqn
corresponding to a clockwise rotation by $ \frac{\pi}{2}$
of the type $A^{(1)}_3$ Dynkin diagram.
The null root in this case is given by
\beqn\label{deA3}
\de=\be_0+\be_1+\be_2+\be_3.
\eeqn
The fundamental weights
$h_1,h_2, h_3$ of $\widetilde{W}\bigl(A^{(1)}_3\bigr)$ defined by Equation \eqref{w},
are related to coroots $\oc\be_1, \oc\be_2, \oc\be_3$
by Cartan matrix of type $A_3$,
\beqn
\bp
\oc\be_1\\
\oc\be_2\\
\oc\be_3
\ep=\bp
2&-1&0\\
-1&2&-1\\
0&-1&2
\ep\bp
h_1\\
h_2\\
h_3
\ep.
\eeqn
Translation of $h_1$ is given by the element  
\beqn\label{th1A30}
t_{h_1}=p_1p_2s_{\be_3}s_{\be_2}s_{\be_1},
\eeqn
which as matrices of linear transforms in 
bases $\{h_1,h_2, h_3,h_{\de}\}\subset V^{(1)\ast}$ 
and $\{\be_1, \be_2, \be_3, \de\}\subset V^{(1)}$ are respectively:
\beqn
t_{h_1}^{h}=
\bp
1&0&1\\
0&1&0\\
0&0&1
\ep\quad\mbox{and}\quad
t_{h_1}^{\de}=
\bp
1&0&-1\\
0&1&0\\
0&0&1
\ep.
\eeqn
From matrix $t_{h_1}^{\de}$, expression of $\de$ given in Equation \eqref{deA3} and condition given by Equation \eqref{Tranc} we see that 
action of $t_{h_1}$ on the $A^{(1)}_3$ type simple system is given by
\beqn\label{th1A3a}
t_{h_1} : \{\be_0, \be_1, \be_2, \be_3\} \mapsto\{\be_0+\de, \be_1-\de, \be_2, \be_3\}.
\eeqn

The weight lattice $P=\mathbb Z \{ h_1, h_2, h_3\}$ of $\widetilde{W}\bigl(A^{(1)}_3\bigr)$ is three dimensional.
In order to describe translation in a general direction on the weight lattice
we need two more linearly independent directions.
Using Equation \eqref{cj}, we can define
\beqn
t_{i}=
(p_1p_2)^{i-1}t_{h1}(p_1p_2)^{1-i}\quad \mbox{for}\quad i\in \mathbb{Z}/4\mathbb{Z}.
\eeqn
That is we have,
\begin{align}\nonumber
t_1&=p_1p_2s_{\be_3}s_{\be_2}s_{\be_1}=t_{h_1},\\
t_2&=(p_1p_2)t_{h_1}(p_1p_2)^{-1}=p_1p_2s_{\be_0}s_{\be_3}s_{\be_1}=t_{h_2-h_1},\\
t_3&=(p_1p_2)^2t_{h_1}(p_1p_2)^{-2}=p_1p_2s_{\be_1}s_{\be_0}s_{\be_3}=t_{h_3-h_2},\\
t_4&=(p_1p_2)^3t_{h_1}(p_1p_2)^{-3}=p_1p_2s_{\be_2}s_{\be_1}s_{\be_0}=t_{h_3}^{-1}=t_{-h_3},
\end{align}
where subscript in the lower case $t$ notations indicates the direction of translation on $P$ in terms of 
the three fundamental weight vectors.
Notice that we have $t_1t_2t_3t_4=1$, so essentially there are only three independent directions.
}

\subsection{Centralizer and normalizer}\label{SCN}
Two objects relevant for our later discussion are centralizer and normalizer of a subgroup of a Coxeter group. For this purpose,
let us equip the real vector space $V^{(1)}$ with a degenerate bilinear
form $( {}\,, {} ):$ $V^{(1)}\times V^{(1)} \to \mb R$,
given by
\beqn\label{Aaij}
A_{ji}=2\frac{(\al_i,\al_j)}{(\al_i, \al_i)},
\eeqn 
where $(\al_i, \al_i)$ gives the squared length of the root $\al_i$.
Note that we have
\beqn\label{aijm}
(\al_i,\al_i)A_{ji}=A_{ij}(\al_j, \al_j).
\eeqn
For simply-laced types we have $A_{ji}=A_{ij}=-1$, that is all the simple roots are of the same length which we set to be $2$ without any lost of generality. 

In the case of non-simply-laced types, we have
two different root length given by
\beqn
\frac{A_{ij}}{A_{ji}}=\frac{(\al_i,\al_i)}{(\al_j,\al_j)}=2,\quad\mbox{or}\quad 3
\eeqn
 for the case of $m_{ij}=4$ or $6$ respectively, where we have let $\al_i$ with $(\al_i, \al_i)=2$, to be the long root and $\al_j$ the short root.
The bilinear
form is degenerate since by definition of $\de$ and the generalised Cartan matrix we have $(\de\,,\al_i)=0$ for all $i\in \{0,1, ..., n\}$.
In particular,  we have $(\de\,,\de)=0$.

In terms of the bilinear form $( {}\,, {} )$, action of the reflection element 
$s_{\al}\in \widetilde{W}^{(1)}$
for $\al\in\Phi^{(1)}$,
is given by 
\beqn\label{refE}
s_{\al}(v)=v-\frac{2(\al\,, v)}{(\al, \al)}\al, 
\eeqn
for any $v\in V^{(1)}$.
We see that
$s_{\al}$ fixes $v$ whenever  $(\al\,, v)=0$. Hence, 
$s_{\al}$ acts trivially on $\de$ for all $\al\in\Phi^{(1)}$.

Let $\be$ be any root of $\Phi^{(1)}$. 
{\it The centralizer} of $\be$ denoted by $C(\be)$ is a reflection subgroup of $W^{(1)}$,
generated by $\lan s_{\al}\mid \al\in \Phi^{(1)}
\; \mbox{and}\; ( \al,\be)=0\ran$. 
We can chose $\be=\al_0$ without any lost of generality since 
$W^{(1)}$ acts transitively on $\Phi^{(1)}$ and look at $C(\al_0)=\lan s_{\al}\mid \al\in \Phi^{(1)}
\; \mbox{and}\; (\al,\al_0)=0\ran$. 

Let $J\subset \De^{(1)}$, the group
 $W_J=\langle s_i\mid \al_i\in J\rangle$ 
is called {\it standard parabolic subgroup} of $\widetilde{W}^{(1)}$, and its conjugates are 
{\it parabolic subgroups}.
{\it The normaliser} of $W_J$ in $\widetilde{W}^{(1)}$ is defined by 

\begin{align}
N(W_J)&=\{g\in \widetilde{W}^{(1)}\mid g^{-1}W_Jg=W_J\}\\\nonumber
&=N_J\ltimes W_J,
\end{align}
which says that the normalizer is the semidirect product of $W_J$ by a complement $N_J$.
 In particular, we have
\beqn\label{nJ}
N_J=\{w\in \widetilde{W}^{(1)}\mid wJ=J\}.
\eeqn 
That is, element of $N_J$
either fixes or permutes the elements of $J$. In other word $N_J$ is the setwise stabilizer of $J$.
{\Rem For an arbitraty group $G$, the normaliser of a subgroup usually
is no more than the subgroup itself. However, it was shown by
Brink and Howlett \cite{BH} that when $G$ is a Coxeter group and the subgroup a parabolic one, 
one can have interesting and highly non-trivial $N_J$. In \cite{BH}, a systematic 
description and an algorithm to
calculate normalizers of parabolic subgroups for an arbitrary Coxeter group
was introduced.}

In the present work, we are
interested in the normalizer of $W_{J'}=\langle s_i\mid \al_i\in J'\rangle$, the affine extension of a parabolic subgroup $W_{J}$, that is
\beqn
J'=J\cup\de-\widetilde{\al}_{J},
\eeqn
where $J\subset \De^{(1)}$ and $\tilde{\al}_{J}$ is the highest root of the finite root system generated by $J$. 
That is we want to investigate 
\begin{align}\label{Nor1}
N(W_{J'})&=\{g\in \widetilde{W}^{(1)}\mid g^{-1}W_{J'}g=W_{J'}\}\\\nonumber
&=N_{J'}\ltimes W_{J'}.
\end{align}
The group
$N_{J'}$ is the setwise stabilizer of $J'$, 
consisting of elements that stabilise the set $J$ and those that exchange elements of 
$J$ with the root $\de-\widetilde{\al}_{J}$. The former are just the elements of $N_{J}$
whose description is given in \cite{BH}. What is left to do is to describe the group elements 
of $\widetilde{W}^{(1)}$ which exchange the roots in $J$ with $\de-\widetilde{\al}_{J}$. 
So although
Brink-Howlett theory of normalizers of parabolic subgroups of Coxeter groups do not apply directly to 
the problems considered in this paper, a large part of
the work is done. 
\section{Centralizer and normalizers in $\widetilde{W}\left(D_5^{(1)}\right)$}
\label{SCND5}
\subsection{$(A_{3}\times A_1\times A_1)^{(1)}$ subroot system}\label{a3a1a1sr}
Let us consider the group $\widetilde{W}\left(D_5^{(1)}\right)$ in Example \ref{ED5}.
The centralizer $C(\al_0)$ is a reflection subgroup of $\widetilde{W}(D^{(1)}_5)$,
generated by all the roots in the $D_5^{(1)}$ root system which
are orthogonal to $\al_0$: $\lan s_{\al}\mid \al\in \Phi^{(1)}
\; \mbox{and}\; ( \al,\al_0)=0\ran$. First let us describe such roots
in the finite $D_5$ root system, that is
$\Om=\{\al\in \Phi\mid ( \al, \al_0 )=0\}$. It is easily
seen that $\Om=\Om_1\cup\Om_2$ is a standard parabolic subsystem of $\Phi$ of type $A_1\times A_3$ generated by two disjoint simple systems of types $A_1$ and $A_3$:
$\{\al_1\}\cup \{\al_3, \al_4, \al_5\}$, corresponding to the vertices of the 
$D_5^{(1)}$ Dynkin diagram that are not joint to the one representing $\al_0$ 
(see Figure \ref{D5}).
The root system of $C(\al_0)$ given by
$\Om^{(1)}=\{\al\in \Phi^{(1)}\mid ( \al, \al_0)=0\}$ has the form
\begin{align}\nonumber
\Om^{(1)}&
=\{\al+k \de\;|\; k\in \mathbb{Z}\; \mbox{and}\; \al\in \Om\}\\\label{a0A1A3}
&=
\{\be+m \de\;|\; m\in \mathbb{Z}\; \mbox{and}\; \be\in \Om_1\}
\cup\{\ga+n \de\;|\; n\in \mathbb{Z}\; \mbox{and}\; \ga\in \Om_2\}\\\nonumber
&=\Om_1^{(1)}\cup\Om_2^{(1)}\cong \left(A_1\times A_3\right)^{(1)}.
\end{align}
The generating set of root system $\Om_i^{(1)}$ contains
those of $\Om_i$ and one extra root: $\de-\tilde{\al}_i$, where
$\tilde{\al}_i$ denotes the highest root of the finite root system $\Om_i$.
This ensures that the affine root systems have the
form $\Om_i^{(1)}=
\{\al+k \de\;|\; k\in \mathbb{Z}\; \mbox{and}\; \al\in \Om_i\}$, for $i=1, 2$.
Here we have $\tilde{\al}_1=\al_1$ and $\tilde{\al}_2=\al_{345}$.
Hence $\Om^{(1)}$
is generated by two disjoint simple systems
$\{\al_1, \de-\al_1\}\cup \{\al_3, \al_4, \al_5, \de-\al_{345}\}
=\{\al_1, \al_{0223345}\}\cup \{\al_3, \al_4, \al_5, \al_{01223}\}$.

To match the embedding given in \cite{Takenawa:03},
let $w=s_{132}$, such that we have $w\al_0=\al_{0123}=\ga_0$, and 
\begin{align}\nonumber
w\al_0&=\al_{0123}=\ga_0,\quad
w\al_{1223345}=\al_{2345}=\ga_1,\\\label{nr}
w\al_1&=\al_{23}=\eta_1,\quad
w\al_{0223345}=\al_{012345}=\eta_0,\\\nonumber
w\al_3&=\al_{12}=\be_1,\quad
w\al_4=\al_{34}=\be_2,\\\nonumber
w\al_5&=\al_{35}=\be_0,\quad
w\al_{01223}=\al_{02}=\be_3.
\end{align}
We refer to the root system generated by $\{\be_0, \be_1, \be_2, \be_3\}$, $\{ \ga_0, \ga_1\}$ and 
$\{\eta_0, \eta_1\}$ 
which is of $(A_{3}\times A_1\times A_1)^{(1)}$ type as
$\be-\ga-\eta$
-system. The corresponding Dynkin diagram is given in
Figure \ref{2A1A3}.

%
From Equation \eqref{a0A1A3} we know that the $\left(A_1\times A_3\right)^{(1)}$ type
root system of $C(\ga_0)=C(w\al_0)$ is given by
$w\Om^{(1)}=w\Om_1^{(1)}\cup w\Om_2^{(1)}$, 
generated by 
$\{w\al_1, w\al_{0223345}\}\cup \{w\al_3, w\al_4, w\al_5, w\al_{01223}\} 
=\{\eta_0, \eta_1\}\cup\{\be_0, \be_1, \be_2, \be_3\}$.
The group $C(\ga_0)$ is then given by
\beqn\label{gA1A3}
C(\ga_0)=\lan s_{\eta_0}, s_{\eta_1}\ran\times\lan s_{\be_0}, s_{\be_1}, s_{\be_2}, s_{\be_3}\ran
=W_{\eta}\times W_{\be}\cong {W}\left(A_1\times A_3\right)^{(1)}.
\eeqn
One can easily find the reflection elements that generate the groups $W_\ga$,
$W_\eta$ and $W_\be$
using the fact 
$s_{w(\al)}=ws_\al w^{-1}$, for $\al\in \De^{(1)}$ and $w\in W(D^{(1)}_5)$. We have:
\begin{align}\nonumber
s_{\ga_0}=s_{\al_{0123}}&=s_{302}s_1s_{203},\quad
s_{\ga_1}=s_{\al_{2345}}=s_{253}s_4s_{352},\\
s_{\eta_1}=s_{\al_{23}}&=s_{232},\quad
s_{\eta_0}=s_{\al_{012345}}=s_{0145}s_{232}s_{5410},\\\nonumber
s_{\be_1}=s_{\al_{12}}&=s_{121},\quad
s_{\be_2}=s_{\al_{34}}=s_{343},\\\nonumber
s_{\be_0}=s_{\al_{35}}&=s_{353},\quad
s_{\be_3}=s_{\al_{02}}=s_{020}.
\end{align}
\subsection{The first variation on the $(A_{3}\times A_1\times A_1)^{(1)}$ 
subsystem}\label{Squasi}
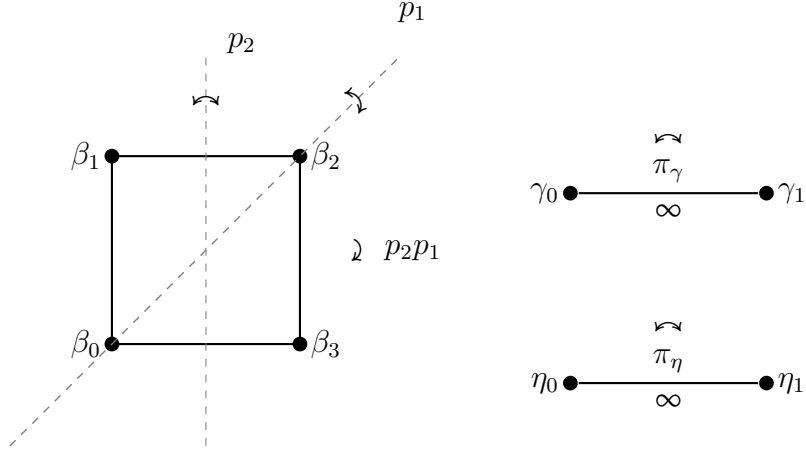
\begin{figure}[t]
\begin{center}
%
\begin{tikzpicture}[scale=.5]
\filldraw [black]  (2,2) node [anchor=east] {$\be_0$} circle(1ex);
\filldraw [black] (2,7) node [anchor=east] {$\be_1$} circle(1ex);
\filldraw [black]  (7,2) node [anchor=west]{$\be_3$} circle(1ex);
\filldraw [black]  (7,7) node [anchor=west] {$\be_{2}$ } circle(1ex);
\draw [black]  (4.5,2) node [anchor=north] {} ;
\draw [black]  (4.5,7) node [anchor=south] {} ;
\draw [black]  (2,4.5) node [anchor=east] {} ;
\draw [black]  (7,4.5) node [anchor=west] { } ;

\draw [thick] (2,7) -- (7,7);
\draw [thick] (7,2) -- (7,7);
\draw [thick](2,2) -- (2,7);
\draw [thick] (2,2) -- (7,2);
\node (sw) at (-1,-1) {}; 
\node (w) at (-1,4.5) {}; 
\node (s) at (4.5,-1) {}; 
\node (n) at (4.5,10) [label=0:$p_{2}$] {}; 
\node (se) at (10,-1) {}; 
\node (ne) at (10, 10) [label=90:$p_{1}$]{}; 
\draw[dashed, gray] (sw)--(ne);
\draw[dashed, gray] (sw)--(ne);
\draw[dashed, gray] (s)--(n);
\draw (8.5, 4.5) node [rotate=-90,label=90:$p_{2}p_{1}$] {$\curvearrowright$};
\draw (8.5, 8.5) node [rotate=-145] {$\curvearrowupdown$};
\draw (4.5, 8.5) node [rotate=-90] {$\curvearrowupdown$};
\end{tikzpicture}
\begin{tikzpicture}[baseline={([yshift={-28ex}]current bounding box.north)},thick]
\node  (a1) {};
\node [right=of a1](a2){};
\node [right=of a2](a3) {};
\node [right=of a3](a4) {};
\node [right=of a4](a5){};
\node [above=of a3, wei](a8) {.};
\node [below=of a3, wei](a9) {.};
\node [above=of a1, wei](a10) {.};
\node [below=of a1, wei](a11) {.};
\node [left=of a1](an){};
\node [above=of a2](a12) {};
\node (r1) at ($(a8)!0.5!(a10)$) {};
\node  [above=of r1] (r3){};
\draw (r1) node [anchor=north] {$\infty$} ;
\draw (r1) node [anchor=south] {$\pi_{\ga}$} ;
\node (r2) at ($(a9)!0.5!(a11)$) {};
\node  [below=of r2] (r4){};
\draw (r2) node [anchor=north] {$\infty$} ;
\draw (r2) node [anchor=south] {$\pi_{\eta}$} ;
\draw (a8) node [anchor=west] {$\ga_1$} ;
\draw (a9) node [anchor=west] {$\eta_1$} ;
\draw (a10) node [anchor=east] {$\ga_0$} ;
\draw (a11) node [anchor=east] {$\eta_0$} ;
\draw (1.3, 2) node  [rotate=-90] {$\curvearrowupdown$};
\draw (1.3, -.5) node  [rotate=-90] {$\curvearrowupdown$};
\draw[-] (a8) --  (a10);
\draw[-] (a9) -- (a11);
\end{tikzpicture}
%
\caption{Dynkin diagram of $(A_{3}\times A_1\times A_1)^{(1)}$
$\ga-\eta-\be$ system}\label{2A1A3}
\end{center}
\end{figure}

Now consider the $ A_1^{(1)}$ type 
Weyl group $W_{\ga}=\lan s_{\ga_0}, s_{\ga_1}\ran$. According to Equation \eqref{Nor1},
normalizer of
$W_{\ga}$ in $\widetilde{W}\left(D_5^{(1)}\right)$ is given by
$N(W_{\ga})=N_{\ga}\ltimes W_{\ga}$, where $N_{\ga}$ is the setwise stabiliser of 
the simple $\ga$-system $\{{\ga_0}, {\ga_1}\}$. 
Centraliser $C(\ga_0)$ fixes point-wisely the simple $\ga$-system $\{{\ga_0}, {\ga_1}\}$.
It can be easily checked that diagram automorphism $\s_1\s_2$ 
exchanges ${\ga_0}$ and ${\ga_1}$, hence is in the normalizer.
Let
the element $g'$ be a minimum length representative in the subset of elements of ${W}\left(D_5^{(1)}\right)$ that exchange ${\ga_0}$ and ${\ga_1}$.
It can be shown that $g'=s_{0145}$. So we have $N_{\ga}=C(\ga_0)\rtimes\lan g', \s_1\s_2\ran$.
Moreover $g'$ commutes
with $\s_1\s_2$ so that the group $\lan g', \s_1\s_2\ran$ is of order
order 8, acting as diagram automorphisms on the simple systems of $\ga$, $\eta$ and $\be$-systems. 
We list actions of some of its elements 
in Equation \eqref{actl}. Element that exchanges ${\ga_0}$ and ${\ga_1}$ is
indicated by having a $\pi_{\ga}$ in the corresponding row. 
Similarly, actions that exchanges the two roots of the $\eta$-system is
indicated by a $\pi_{\eta}$ in the corresponding row. Recall that diagram automorphisms on an $A_3^{(1)}$ $\be$-system (see Figure \ref{2A1A3}) are generated by $p_1$ and $p_2$,
\begin{align*}
p_1&: \{\be_1\leftrightarrow\be_3\},\\
p_2&: \{\be_1\leftrightarrow\be_2,\be_0\leftrightarrow\be_3\}.
\end{align*}
We use these to indicate the actions of elements of $\lan g', \s_1\s_2\ran$
on the $\be$-system.
Whenever an element acts trivially on a system, the symbol $-$ is used.
\beqn\label{actl}
\begin{matrix}
          &\eta &\ga &\be\\
\s_2\s_1:     &-    &\pi_{\ga}  &p_2p_1     \\
\s_1\s_2:     &-    &\pi_{\ga}  &p_1p_2     \\
\s_1\s_2\s_1\s_2:     &-    &-  &p_1p_2p_1p_2     \\
g'=s_{0145}: &\pi_{\eta}    &\pi_{\ga}  &p_1p_2p_1p_2     \\
g'\s_1\s_2\s_1\s_2: &\pi_{\eta}    &\pi_{\ga}  &- \quad\quad .
\end{matrix}
\eeqn
From Equation \eqref{actl}, we can see that the element $g'\s_1\s_2\s_1\s_2$ acts
on $\eta$-$\ga$-systems only, while products of $\s_1$ and $\s_2$ acts on $\be$-$\ga$-systems only. Hence we have the following decomposition of $N_{\ga}$ as a direct products of two extended affine Weyl groups,
\begin{align}\nonumber
N_{\ga}&=C(\ga_0)\rtimes\lan g', \s_1\s_2\ran
=\left(W_{\eta}\times W_{\be}\right)\rtimes\lan g', \s_1\s_2\ran\\\label{n}
&=\lan s_{\eta_0}, s_{\eta_1},g'\s_1\s_2\s_1\s_2\ran\times\lan s_{\be_0}, s_{\be_1}, s_{\be_2}, s_{\be_3},
\s_1\s_2\ran\\\nonumber
&=\widetilde{W}_{\eta}\times \widetilde{W}_{\be}\cong \widetilde{W}\left(A_1\times A_3\right)^{(1)}.
\end{align}
Finally we have the normalizer of $W_{\ga}$ in $\widetilde{W}(D^{(1)}_{5})$:
\begin{align}\label{nA131}
N(W_{\ga})=N_{\ga}\ltimes W_{\ga}
&=\left(\widetilde{W}_{\eta}\times \widetilde{W}_{\be}\right)\ltimes W_{\ga}
\cong\widetilde{W}\left(A_1\times A_3\right)^{(1)}\ltimes{W}(A_1^{(1)})
\end{align}
Given the $A_1^{(1)}$ and $A_3^{(1)}$ type extended affine Weyl groups 
of $\ga$ and $\be$-systems in the normalizer
above one can follow the
recipe in Examples \ref{TranA1} and \ref{TranA3} to construct elements of
quasi-translation.
First, take the $A_1^{(1)}$ type $\eta$-system 
$\widetilde{W}_{\eta}=\lan s_{\eta_0}, s_{\eta_1},g'\s_1\s_2\s_1\s_2\ran$.
By Equation \eqref{th1}, define element $t_{h_{\eta_1}}$ by
\beqn\label{nA313t}
t_{h_{\eta_1}}=g'\s_1\s_2\s_1\s_2s_{\eta_1}=g'\s_1\s_2\s_1\s_2s_{232},
\eeqn
where the subscript indicates its association with the fundamental
weight $h_{\eta_1}$ of $\eta$-system. Action of $t_{h_{\eta_1}}$ on
the simple roots of the $D^{(1)}_{5}$ system is given by,
\begin{align}\nonumber
t_{h_{\eta_1}}&:\{\al_0,\al_1,\al_2,\al_3,\al_4,\al_5\}\\\label{etah1}
&\mapsto \{\al_{23450},\al_{12345},-\al_{345},\al_{012},\al_{01234},\al_{01235}\}.
\end{align}
We can clearly see that $t_{h_{\eta_1}}$ is not an element of translation in
$\widetilde{W}(D^{(1)}_{5})$. This fact is better understood by looking at its action on 
the $\ga-\eta-\be$-system:
\begin{align}\nonumber
t_{h_{\eta_1}}&:=\{ \ga_0, \ga_1,\eta_0, \eta_1,\be_0, \be_1, \be_2, \be_3\}\\\label{e1geb}
&\mapsto \{ \ga_1, \ga_0,\eta_0+\de, \eta_1-\de,\be_0, \be_1, \be_2, \be_3\}.
\end{align}
While $t_{h_{\eta_1}}$ behaves like a translation on the $\eta$ and $\be$-systems, it acts
as a permutation of order 2 on the $\ga$-system. This permutative effect on the $\ga$-system can be  eliminated by iterating it twice, that is
\begin{align}\nonumber
t_{h_{\eta_1}}^2&:\{\al_0,\al_1,\al_2,\al_3,\al_4,\al_5\}\\\label{e3}
&\mapsto \{\al_{0}+\de,\al_{1}+\de,\al_{2}-\de,\al_{3}-\de,\al_{4}+\de,\al_{5}+\de\},
\end{align}
which in fact is a translation in $\widetilde{W}(D^{(1)}_{5})$ by $2h_{\eta_1}=\eta_1$.
We refer to elements such as $t_{h_{\eta_1}}$ quasi-translations.
We do not see this permutative action of $t_{h_{\eta_1}}$ when considering only the subspace of $V$ orthogonal to $\ga$-system. 
That is, we have $V=V_{\ga}\oplus V_{\ga}^{\perp}$, where
$V_{\ga}=\mbox{Span}(\{ \ga_0, \ga_1\})$, and 
$V_{\ga}^{\perp}=\mbox{Span}(\{ \eta_0, \eta_1,\be_0, \be_1, \be_2, \be_3\})$. The element $t_{h_{\eta_1}}$ is then a translation on 
the subspace $V_{\ga}^{\perp}$.

For $A_3^{(1)}$ type $\be$-system $\widetilde{W}_{\be}=\lan s_{\be_0}, s_{\be_1}, s_{\be_2}, s_{\be_3},
\s_1\s_2\ran$,
quasi-translational element 
associated to the fundamental weight $h_{\be_1}$
is given by the formula in Equation \eqref{th1A30},
\beqn\label{th1A3}
t_{h_{\be_1}}=\s_1\s_2s_{\be_3}s_{\be_2}s_{\be_1}
=\s_1\s_2s_{020}s_{343}s_{121}.
\eeqn
Element $t_{h_{\be_1}}$ acts like a translation on the $\be$-system
while permutes the two simple roots of $\ga$-system:
\begin{align}\nonumber
t_{h_{\be_1}}&:=\{ \ga_0, \ga_1,\eta_0, \eta_1,\be_0, \be_1, \be_2, \be_3\}\\\label{hb1}
&\mapsto \{ \ga_1, \ga_0,\eta_0, \eta_1,\be_0+\de, \be_1-\de, \be_2, \be_3\}.
\end{align}
Simple roots of $\be$-system have the form
$\be_j=\sum_{i=1}^{5}a^i_j\al_i$, and the corresponding coroots 
$\oc\be_j=\sum_{i=1}^{5}a^i_j\oc\al_i$ (for $j=1, 2, 3$), where coefficients $a^i_j$
are given in Equation \eqref{nr}.  Fundamental weights
of the $\be$-system are then given by
\beqn\label{brw}
\bp
\oc\be_1\\
\oc\be_2\\
\oc\be_3
\ep=\bp
2&-1&0\\
-1&2&-1\\
0&-1&2
\ep\bp
h_{\be_1}\\
h_{\be_2}\\
h_{\be_3}
\ep.
\eeqn
Quasi-translations along the other directions of weight lattice
of the $\be$-system can be then given by 
\[
T_{i}=
(\s_1\s_2)^{i-1}t_{h_{\be_1}}(\s_1\s_2)^{1-i}\quad \mbox{for}\quad i\in \mathbb{Z}/4\mathbb{Z}.
\]
That is we have,
\begin{align}\label{tb1}
T_1&=\s_1\s_2s_{\be_3}s_{\be_2}s_{\be_1}=t_{h_{\be_1}},\\\label{tb2}
T_2&=(\s_1\s_2)t_{h_{\be_1}}(\s_1\s_2)^{-1}=\s_1\s_2s_{\be_0}s_{\be_3}s_{\be_1}=t_{h_{\be_2}-h_{\be_1}},\\\label{tb3}
T_3&=(\s_1\s_2)^2t_{h_{\be_1}}(\s_1\s_2)^{-2}=\s_1\s_2s_{\be_1}s_{\be_0}s_{\be_3}=t_{h_{\be_3}-h_{\be_2}},\\\label{tb4}
T_4&=(\s_1\s_2)^3t_{h_{\be_1}}(\s_1\s_2)^{-3}=\s_1\s_2s_{\be_2}s_{\be_1}s_{\be_0}=t_{h_{\be_3}}^{-1}=t_{-h_{\be_3}},
\end{align}
where $T_1T_2T_3T_4=1$.

\subsection{A second variation on the $(A_{3}\times A_1\times A_1)^{(1)}$ 
subsystem}\label{2var}
For the same underlying $(A_{3}\times A_1\times A_1)^{(1)}$ type $\be-\eta-\ga$ 
subroot system
another subgroup $\widetilde{W}\left(A_1\times A_1\right)^{(1)}\ltimes {W}(A_3^{(1)})$ of $\widetilde{W}(D^{(1)}_{5})$ can be formed by considering the 
normalizer of the $A_3^{(1)}$ type $\be$-system
$W_\be=\lan s_{\be_0}, s_{\be_1}, s_{\be_2}, s_{\be_3}\ran$ in $\widetilde{W}(D_5^{(1)})$.
We have $N(W_{\be})=N_{\be}\ltimes W_{\be}$, where
$N_{\be}=C(\be)\rtimes\lan g', \s_1\s_2\ran$ is the setwise stabilizer of $\{\be_1, \be_2, \be_0, \be_3\}$. The centralizer of $\be$-system
$C(\be)$ is generated by reflections along roots of the $D_5^{(1)}$ root system which are orthogonal to the $\be$-system. From our earlier discussion in Section \ref{a3a1a1sr} we know they form the $\eta$ and $\ga$-systems given in Equation \eqref{nr}. Hence we have
\beqn\label{A1A1}
C(\be)=W_\eta\times W_\ga=\lan s_{\eta_0}, s_{\eta_1}\ran\times\lan s_{\ga_0}, s_{\ga_1}\ran\cong{W}\left(A_1\times A_1\right)^{(1)}.
\eeqn
The $\lan g', \s_1\s_2\ran$ part of $N_{\be}$ permutes the simple $\be$-system, acting on it as diagram automorphisms.
Moreover, it can be decomposed 
into elements that either act on $\eta-\be$-system or 
$\ga-\be$-system only, so that a direct product of two extended Weyl groups 
for the $\eta$ and $\ga$-systems given in Equation \eqref{A1A1} can be formed.
An inspection of Equation \eqref{actl} tells us that $\s_1\s_2$, and $g'\s_1\s_2$
are just such two elements. Hence normalizer of the $W_{\be}$ in $\widetilde{W}(D_5^{(1)})$ is given by
\begin{align}\label{a1a1a3}
N(W_{\be})=N_{\be}\ltimes W_{\be}
&=\left(C(\be)\rtimes\lan g', \s_1\s_2\ran\right)\ltimes\lan s_{\be_0}, s_{\be_1}, s_{\be_2}, s_{\be_3}\ran\\\nonumber
&=\left(\lan s_{\eta_0}, s_{\eta_1},g'\s_1\s_2\ran\times\lan s_{\ga_0}, s_{\ga_1},\s_1\s_2\ran\right)
\ltimes\lan s_{\be_0}, s_{\be_1}, s_{\be_2}, s_{\be_3}\ran\\\nonumber
&=\left(\widetilde{W}_\eta\times \widetilde{W}_\ga\right)\ltimes W_{\be}\\\nonumber
&\cong \widetilde{W}\left(A_1\times A_1\right)^{(1)}\ltimes W(A_3^{(1)}).
\end{align}

Two quasi-translations can be defined for the
$\widetilde{W}\left(A_1^{(1)}\right)$ type $\eta$ and $\ga$-systems in the normalizer above,
associated with fundamental weights $h_{\eta_1}$ and $h_{\ga_1}$, respectively.
We have
\beqn
t_{h_{\eta_1}}=g'\s_1\s_2s_{\eta_1}=g'\s_1\s_2s_{232},\quad\mbox{and}\quad
\quad t_{h_{\ga_1}}=\s_1\s_2s_{\ga_1}=\s_1\s_2s_{2534352}.
\eeqn
Their actions on $\ga-\eta-\be$-system are given respectively by
\begin{align}\label{te1}
t_{h_{\eta_1}}&:=\{ \ga_0, \ga_1,\eta_0, \eta_1,\be_0, \be_1, \be_2, \be_3\}\\\nonumber
&\mapsto \{ \ga_0, \ga_1,\eta_0+\de, \eta_1-\de,\be_3, \be_0, \be_1, \be_2\},
\end{align}
and
\begin{align}\label{tg1}
t_{h_{\ga_1}}&:=\{ \ga_0, \ga_1,\eta_0, \eta_1,\be_0, \be_1, \be_2, \be_3\}\\\nonumber
&\mapsto \{ \ga_0+\de, \ga_1-\de,\eta_0, \eta_1,\be_1, \be_2, \be_3, \be_0\}.
\end{align}
Elements
$t_{h_{\ga_1}}$ and $t_{h_{\eta_1}}$ act on $\be$-system as permutations of order 4, thus they
become translations
after four iterations, that is $t_{h_{\ga_1}}^4$ and $t_{h_{\eta_1}}^4$
are translations on the $D_5$ weight lattice. 
\section{Normalizer theory in the context of a $\widetilde{W}(D_5^{(1)})$ $q$-\Pa equation}\label{TOS}
\subsection{Takenawa's embedding of KNY's $\widetilde{W}\left(A_1\times A_3\right)^{(1)}$
system}
We see that it corresponds exactly to 
the normalizer 
$\widetilde{W}\left(A_1\times A_3\right)^{(1)}\ltimes{W}(A_1^{(1)})$
in $\widetilde{W}(D_5^{(1)})$ given
in Equation \eqref{nA131}, where we have chosen a 
conjugation of the $(A_{3}\times A_1\times A_1)^{(1)}$ type subroot system of 
$\widetilde{W}(D_5^{(1)})$ to coincide with that of \cite{Takenawa:03}. The element
$\phi$ which gives rise to discrete \Pa equation \cite[Thm 3.2]{Takenawa:03}
corresponds to $t_{h_{\eta_1}}$ given by Equation \eqref{nA313t}, an element of 
infinite order in the $N_\ga$ part of the normalizer, 
which acts permutatively on the simple $\ga$-system given in Equation \eqref{e1geb}. 
The fact that
$\phi$ iterated twice becomes a translation is due to the fact that it acts on  
$\ga$-system as a permutation of order 2. The $\widetilde{W}\left(A_1\times A_3\right)^{(1)}$ symmetry of the KNY system is recovered by considering
only the subspace orthogonal to the $\ga$-system, which is spanned by 
the $\left(A_1\times A_3\right)^{(1)}$ type
$\eta-\be$-system.

\subsection{Okubo-Suzuki system $n=1$ case, $(A_{3}\times A_1\times A_1)^{(1)}$}
In 2018, Okubo and Suzuki proposed a new $(A_{2n+1}\times A_1\times A_1)^{(1)}$ type
generalisation of Sakai's $\widetilde{W}(D_5^{(1)})$ $q$-PVI equation 
from the framework of Cluster algebra, which we refer as the OS-system. It contains four 
previously known generalisations of the $q$-PVI equation \cite{OS:18}:
\bit
\item $T_1$: $q$-Drinfeld-Sokolov system $q$-$P_{(n+1, n+1)}$ \cite{Suz:17},
\item $T_2$: Sakai's $q$-Garnier system \cite{Sak:05},
\item $T_3$: Nagao-Yamada's variation of the $q$-Garnier system \cite{NY:18},
\item $T_4$: Tsuda's $q$-UC hierarchy \cite{Tsuda:10}.
\eit
However, if one looks at the defining relations for the Weyl group symmetries of 
OS system
given in \cite [Thm 2.1]{OS:18}, it can be seen
that we have in fact the Weyl group $\widetilde{W}\left(A_1\times A_1\right)^{(1)}\ltimes{W}(A_{2n+1}^{(1)})$. 
For the case $n=1$, the $\widetilde{W}\left(A_1\times A_1\right)^{(1)}\ltimes{W}(A_{3}^{(1)})$ type system, as we have shown in Section \ref{2var},
can be embedded inside a $\widetilde{W}(D_5^{(1)})$ symmetry group. 
In fact, we have shown that it arises exactly as the normalizer of $A_3^{(1)}$ type $\be$-system
$W_\be=\lan s_{\be_0}, s_{\be_1}, s_{\be_2}, s_{\be_3}\ran$ in $\widetilde{W}(D_5^{(1)})$.

Now we are in the position to analyse in detail 
the nature of the four directions $T_i$ $(1\leq i \leq 4)$  
in terms of the properties of the two
normalizers 
$\widetilde{W}\left(A_1\times A_1\right)^{(1)}\ltimes{W}(A_3^{(1)})$ and
$\widetilde{W}\left(A_1\times A_3\right)^{(1)}\ltimes{W}(A_1^{(1)})$ 
with an underlying subroot system $(A_{3}\times A_1\times A_1)^{(1)}$ of $\widetilde{W}(D_5^{(1)})$. In particular, we describe the
four directions in terms of the fundamental weights $h_i$ ($ 1\leq i \leq 5$)
of $\widetilde{W}(D_5^{(1)})$,
given by Equation \eqref{wD5}.

The correspondence between our notations and 
those in \cite{OS:18} is given as follows,
\begin{align}\nonumber
s_{\eta_1}&=s_1,\quad s_{\eta_0}=s_0, \quad g'\s_1\s_2=\pi,\\\label{notrel}
s_{\ga_1}&=s_1',\quad s_{\ga_0}=s_0', \quad \s_2\s_1=\pi',\\\nonumber
s_{\be_1}&=r_1,\quad s_{\be_2}=r_2,\quad s_{\be_3}=r_3,\quad s_{\be_0}=r_0.
\end{align}
The four directions $T_i$ ($1\leq i\leq 4$) given in \cite{OS:18}
are then given by,
\begin{align}
T_1&=s_{\ga_1}\s_2\s_1s_{\eta_1}(g'\s_2\s_1)^{-1},\\
T_2&=\left(s_{\be_0}s_{\be_1}s_{\be_2}(\s_2\s_1)^{-1}\right)^2,\\
T_3&=s_{\ga_1}s_{\be_1}s_{\be_2}s_{\be_3}(\s_2\s_1)^{-1},\\
T_4&=\left(s_{\be_0}s_{\be_2}(\s_2\s_1)^{-1}\right)^2.
\end{align}

First, let us look at the $T_1$ direction. We have
\begin{align}\nonumber
T_1&=s_{\ga_1}\s_2\s_1s_{\eta_1}(g'\s_1\s_2)^{-1}\\\nonumber
&=s_{\ga_1}(\s_1\s_2)^{-1}s_{\eta_1}(g'\s_1\s_2)^{-1}\\\nonumber
&=(\s_1\s_2s_{\ga_1})^{-1}(g'\s_1\s_2s_{\eta_1})^{-1}\\
&=(t_{h_{\ga_1}}t_{h_{\eta_1}})^{-1}.
\end{align}

In general, elements such as $t_{h_{\ga_1}}$ or $t_{h_{\eta_1}}$ 
are not translations, since
they act on $\be$-system as permutations of order four as shown in Equations
\eqref{te1} and \eqref{tg1}. However, composition $t_{h_{\ga_1}}t_{h{\eta_1}}$
is a translation since the permutative actions of $t_{h_{\ga_1}}$ and $t_{h_{\eta_1}}$
on $\be$-system are exactly
inverse of each other so applying one after another cancel out the permutative effect and we
obtain indeed a translation on $\ga-\eta-\be$-system,
\begin{align}\nonumber
(t_{\ga_1}t_{\eta_1})^{-1}&:=\{ \ga_0, \ga_1,\eta_0, \eta_1,\be_0, \be_1, \be_2, \be_3\}\\\label{T11}
&\mapsto \{ \ga_0-\de, \ga_1+\de,\eta_0-\de, \eta_1+\de,\be_0, \be_1, \be_2, \be_3\},
\end{align}
or on the simple system of $D_5^{(1)}$,
\begin{align}\nonumber
(t_{\ga_1}t_{\eta_1})^{-1}&:\{\al_0,\al_1,\al_2,\al_3,\al_4,\al_5\}\\\label{T12}
&\mapsto \{\al_{0}-\de,\al_{1}-\de,\al_2+\de,\al_3,\al_4,\al_5\}.
\end{align}
We see that the element $(t_{\ga_1}t_{\eta_1})^{-1}$ is
a translation by 
\beqn
-(h_{\ga_1}+h_{\eta_1})=-\frac{\oc\ga_1+\oc\eta_1}{2}=-\frac{\oc\al_{2345}+\oc\al_{23}}{2}
=h_1-h_2.
\eeqn

To describe $T_i$ ($i=2,3,4$) we make use of elements of normalizer
$\left(\widetilde{W}(A_1^{(1)})\times \widetilde{W}(A_3^{(1)})\right)\ltimes{W}(A_1^{(1)})$ given in Equation \eqref{nA131}. In particular, the quasi-translational elements in
$\widetilde{W}(A_3^{(1)})$ part of the normalizer. We have
\begin{align*}
T_2&=\left(s_{\be_0}s_{\be_1} s_{\be_2}\s_2\s_1\right)^2\\
&=\left(\s_1\s_2s_{\be_2}s_{\be_1}s_{\be_0}\right)^{-2},\\
&=(t_{-h_{\be_3}})^{-2},\\
&=t_{2h_{\be_3}}.
\end{align*}
Quasi-translational element $t_{-h_{\be_3}}$, defined in Equation \eqref{tb4}, acts on the ${W}(A_1^{(1)})$ type $\ga$-system
as a permutation of order 2. Therefore, applied twice we have a translation on $\ga-\eta-\be$-system,
\begin{align}\nonumber
t_{2h_{\be_3}}&:=\{ \ga_0, \ga_1,\eta_0, \eta_1,\be_0, \be_1, \be_2, \be_3\}\\\label{T21}
&\mapsto \{ \ga_0, \ga_1,\eta_0, \eta_1+\de,\be_0+2\de, \be_1, \be_2, \be_3-2\de\},
\end{align}
or equivalently on the simple system of $D_5^{(1)}$,
\begin{align}\nonumber
t_{2h_{\be_3}}&:\{\al_0,\al_1,\al_2,\al_3,\al_4,\al_5\}\\\label{T22}
&\mapsto \{\al_{0}-\de,\al_{1}+\de,\al_2-\de,\al_3+\de,\al_4-\de,\al_5+\de\}.
\end{align}
In particular, 
we have 
\beqn 
2h_{\be_3}=(\oc\be_1+2\oc\be_2+3\oc\be_3)/2=(\oc\al_{12}+2\oc\al_{34}+3\oc\al_{02})/2=-h_1+h_2-h_3-h_4+h_5,
\eeqn
where we have used Equations \eqref{brw} and \eqref{rw}.

For $T_3$ direction we have
\begin{align*}
T_3&=s_{\ga_1}s_{\be_1}s_{\be_2}s_{\be_3}(\s_2\s_1)^{-1}\\
&=s_{\ga_1}\left(\s_1\s_2s_{\be_3}s_{\be_2}s_{\be_1}\right)^{-1}\\
&=s_{\ga_1}t_{-h_{\be_1}}.
\end{align*}
Quasi-translational element $t_{h_{\be_1}}$ is defined in Equation \eqref{th1A3},
whose action on the $\ga-\eta-\be$-system is given by Equation \eqref{hb1}.

On composition with $s_{\ga_1}$, which acts on the $\ga-\eta-\be$-system by 
\begin{align}\nonumber
s_{\ga_1}&:=\{ \ga_0, \ga_1,\eta_0, \eta_1,\be_0, \be_1, \be_2, \be_3\}\\
&\mapsto \{ \ga_0+2\ga_1, -\ga_1,\eta_0, \eta_1,\be_0, \be_1, \be_2, \be_3\},
\end{align}
we obtain a translation on the $\ga-\eta-\be$-system
\begin{align}\nonumber
s_{\ga_1}t_{-h_{\be_1}}&:=\{ \ga_0, \ga_1,\eta_0, \eta_1,\be_0, \be_1, \be_2, \be_3\}\\
&\mapsto \{ \ga_0-\de, \ga_1+\de,\eta_0, \eta_1,\be_0-\de, \be_1+\de, \be_2, \be_3\},
\end{align}
or equivalently on the simple system of $D_5^{(1)}$ as
\begin{align}\nonumber
s_{\ga_1}t_{-h_{\be_1}}&:\{\al_0,\al_1,\al_2,\al_3,\al_4,\al_5\}\\\label{e2}
&\mapsto \{\al_0-\de,\al_1,\al_2+\de,\al_3-\de,\al_4+\de,\al_5\}.
\end{align}
The element $s_{\ga_1}t_{-h_{\be_1}}$ is a translation by 
\beqn
-h_{\ga_1}-h_{\be_1}=-\frac{\oc\ga_1}{2}-\frac{3\oc\be_1+2\oc\be_2+\oc\be_3}{4}=-\frac{2\oc\al_1+4\oc\al_2+2\oc\al_3+3\oc\al_4+\oc\al_5}{4}
=-h_2+h_3-h_4.
\eeqn
Lastly for $T_4$ direction we have
\begin{align}\nonumber
T_4&=\left(s_{\be_0}s_{\be_2}(\s_2\s_1)^{-1}\right)^2\\\nonumber
&=s_{\be_0}s_{\be_2}(\s_2\s_1)^{-1}s_{\be_0}s_{\be_2}(\s_2\s_1)^{-1}\\\label{T4}
&=s_{\be_0}s_{\be_2}s_{\be_3}s_{\be_1}(\s_2\s_1)^{-2}\\\nonumber
&=s_{\be_0}s_{\be_3}s_{\be_2}s_{\be_3}s_{\be_2}s_{\be_1}(\s_2\s_1)^{2}\\\nonumber
&=\s_2\s_1s_{\be_3}s_{\be_2}s_{\be_1}\s_2\s_1s_{\be_1}s_{\be_0}s_{\be_3}\\\nonumber
&=t_{h_{\be_1}}t_{h_{\be_3}-h_{\be_2}}.
\end{align}

Since both elements $t_{h_{\be_1}}$ and $t_{h_{\be_3}-h_{\be_2}}$, given
by Equations \eqref{th1A3} and \eqref{tb3} respectively,  permute the simple $\ga$-system: 
$\ga_0\leftrightarrow\ga_1$, applying one after the other cancels out the effect of the permutation thus gives us a translation. Its action on the $\ga-\eta-\be$-system
is given by,
\begin{align}\nonumber
t_{h_{\be_1}}t_{h_{\be_3}-h_{\be_2}}&:=\{ \ga_0, \ga_1,\eta_0, \eta_1,\be_0, \be_1, \be_2, \be_3\}\\
&\mapsto \{ \ga_0, \ga_1,\eta_0, \eta_1,\be_0+\de, \be_1-\de, \be_2+\de, \be_3-\de\},
\end{align}
and on the simple system of $D_5^{(1)}$,
\begin{align}\nonumber
t_{h_{\be_1}}t_{h_{\be_3}-h_{\be_2}}&:\{\al_0,\al_1,\al_2,\al_3,\al_4,\al_5\}\\\label{e2a}
&\mapsto \{\al_0,\al_1,\al_2-\de,\al_3+\de,\al_4,\al_5\}.
\end{align}
Element $t_{h_{\be_1}}t_{h_{\be_3}-h_{\be_2}}$ is a translation on
the $D_5^{(1)}$ weight lattice by
\beqn
h_{\be_1}-h_{\be_2}+h_{\be_3}=\frac{\oc\be_1+\oc\be_3}{2}
=-\oc\al_3-\frac{\oc\al_4}{2}-\frac{\oc\al_5}{2}=h_2-h_3.
\eeqn
{\Rem In general, from a $(D_{2n+1}\times A_1\times A_1)^{(1)}$ subsystem 
of a $D_{2n+3}^{(1)}$ root system, by consider
either the normalizer of a $D_{2n+1}^{(1)}$ or an $A_1^{(1)}$ type 
subsystem two subgroups
$\widetilde{W}\left(A_1\times A_1\right)^{(1)}\ltimes {W}(D_{2n+1}^{(1)})$
and
$\widetilde{W}\left(D_{2n+1}\times A_1\right)^{(1)}\ltimes {W}(A_{1}^{(1)})$
can be respectively formed. 
For the $n=1$ case these two symmetry groups coincide with the OS-system and
Takenawa's embedding in Sakai's $\widetilde{W}\left(D_{5}^{(1)}\right)$ $q$-PVI equation. The normalizer procedure can be used to obtain generalisations of \Pa equations as subsystems systems with
bigger symmetry groups. Such a system
was in fact given by Masuda in his $D_n^{(1)}$ generalisation of Sakai's $\widetilde{W}\left(D_{5}^{(1)}\right)$ $q$-PVI equation \cite{Masuda:15}.
}
\section{Conclusion}\label{Dis}
In the present work, we explained the nature and properties of quasi-translational
elements arising in the study of \Pa equations based on theory of normalizers
of Coxeter groups. 
In particular, we showed that the
quasi-translational nature of such
elements resulted from the fact that
they act permutatively
on some reflection subgroups of the original group.
The order of permutation decides the number of iteration for which they become
actual translations in the weight lattice.
Quasi-translations arise under different guises in a variety of contexts 
in the theory of discrete integrable equations. For example, in ``symmetrisation'' procedures of asymmetric
discrete \Pa equations known as {\it projective reductions}
 \cite{KNT:11, Stokes:18}; from reductions of partial difference equations 
 \cite{ahjn:16}; or as elements that give rise to discrete equations which govern evolutions of
Schramm's circle patterns \cite{cp:17}, 
these we plan to discuss in subsequent publications. 

\bibliographystyle{abbrv}
\bibliography{wn}
\end{document}